\documentclass[final,twocolumn,prl,aps,floats,amsfonts,amssymb,showpacs,superscriptaddress]{revtex4}
\usepackage{graphicx}
\usepackage{color}
\usepackage{psfrag}
\usepackage{gensymb}
\usepackage{enumerate}
\usepackage{amssymb}
\usepackage{amsmath}
\usepackage{wasysym}
\usepackage{hyperref}
\usepackage{color}

\begin{document}

\title{The role of boundary conditions on helicoidal flow collimation: \\ consequences for the Von-K\'{a}rm\'{a}n-Sodium dynamo experiment}

\author{J. Varela}
\affiliation{AIM, CEA/CNRS/University of Paris 7, CEA-Saclay, 91191 Gif-sur-Yvette, France}
\affiliation{LIMSI, CNRS-UPR 3251, Rue John von Neumann, 91405 Orsay, France}
\author{S. Brun}
\affiliation{AIM, CEA/CNRS/University of Paris 7, CEA-Saclay, 91191 Gif-sur-Yvette, France}
\author{B. Dubrulle}
\affiliation{SPEC/IRAMIS/DSM, CEA, CNRS, University Paris-Saclay, CEA Saclay, 91191 Gif-sur-Yvette, France}
\author{C. Nore}
\affiliation{LIMSI, CNRS-UPR 3251, Rue John von Neumann, 91405 Orsay, France}

\date{\today}


\begin{abstract}
We present hydrodynamic and magneto-hydrodynamic simulations of liquid sodium flow with the PLUTO compressible MHD code to investigate influence of magnetic boundary conditions on the collimation of helicoidal motions. We use a simplified cartesian geometry to represent the flow dynamics in the vicinity of one cavity of a multi-blades impeller inspired by those used in the Von-K\'{a}rm\'{a}n-Sodium (VKS) experiment. We show that the impinging of the large scale flow upon the impeller generates a coherent helicoidal vortex inside the blades, located at a distance from the upstream blade piloted by the incident angle of the flow. This vortex collimates any existing magnetic field lines leading to an enhancement of the radial magnetic field that is stronger for ferromagnetic than for conducting blades. The induced magnetic field modifies locally the velocity fluctuations, resulting in an enhanced helicity. This process possibly explains why dynamo action is more easily triggered in the VKS experiment when using soft iron impellers. 
\end{abstract}

\maketitle

\paragraph*{Introduction}
The dynamo effect is the main driver of magnetic field in celestial bodies through the transformation of mechanical energy in magnetic energy \cite{1,2,3}.  Given their huge dimensions and the extreme conditions prevailing in their environment, natural dynamos are characterized by control parameters that are presently out-of-reach of analytical or numerical grasp. To make progress, several liquid metal dynamo experiments have been built \cite{4,5,6}, allowing observation and study of several regimes of astrophysical or geophysical interest with interesting open questions regarding the {\sl dynamo mechanisms}. When the level of fluctuations is low, like in the Riga or Karlsruhe dynamo, the dynamo mechanism appears well described by a "laminar" mechanism, by which the mean velocity flow stretchs and folds the magnetic field lines, resulting in magnetic field growth\cite{2}. The observed magnetic field is then similar to that predicted by kinematic computations. When the level of fluctuations is large, like in the VKS dynamo, the observed dynamo magnetic field is dipolar aligned with the symmetry axis of the set-up, in contradiction with the "laminar" mechanism that predicts an equatorial dipole  \cite{Rav05}. It has therefore been suggested that the dominant mechanism in VKS is, like in many celestial bodies,  "turbulent" \cite{mordant,gissinger,13}, with a  magnetic field generation occurring either via the interaction of the differential rotation and the non-axisymmetric velocity perturbations \cite{mordant}, or via self-interaction of the helical perturbations\cite{7}. However, many features of the VKS dynamo are left unexplained.
The VKS experiment generates turbulence through the counter-rotation of two impellers ({\it{i.e.}} disks fitted with blades). Both the geometry and the composition of the impellers influence the dynamo: i) when the blades curvature is decreased, the dynamo threshold increases \cite{8}. ii) When the blades and the disks are made of soft-iron, with large magnetic permeability, the dynamo threshold is at least 2 to 3 times smaller than when the blades and/or the disks are made of copper, with large electrical conductivity, or with stainless steel \cite{9}. iii) When the ferromagnetic impellers are magnetized, the dynamo threshold decreases \cite{8}. This shows that a better understanding of the VKS dynamo requires investigation of  the complex interaction of the large scale flow, generated by the impeller, with magnetic field and current at the impeller. Previous investigations of the role of impeller material on the dynamo mechanism in VKS were restricted to the role of the boundary conditions onto the magnetic field growth with {\sl prescribed} velocity \cite{13,14,15}. 

In this letter, we study the influence of  a {\sl prescribed} magnetic field (presumably issued from the disk magnetization) on the velocity fluctuations. We use Magneto-HydroDynamic (MHD) numerical simulations in a simplified geometry mimicking the flow structure in the vicinity of the impeller.  We show that the interplay between the magnetic and velocity fields results in a strong enhancement of the helicity only when ferromagnetic boundary conditions are used.


\paragraph*{Numerical model}

We use the PLUTO code with a viscous and resistive MHD single fluid model in 3D Cartesian coordinates \cite{10}. The simulation domain is plotted in Fig. 1A and focused on the near impeller region and the flow in between two blades, with X , Y and Z directions corresponding to local azimuthal (toroidal), radial and vertical (poloidal) directions. For simplicity, we consider straight blades instead of curved blades and walls without thickness. The gray surfaces on Fig. 1A represent the blades (at X$=0$ and X$=2$), the impeller disk (at Z$=0$) and the cylinder outer wall (at Y$=4$). The influence of the blade's geometry is taken into account via the velocity boundary condition, through $\Gamma$, the ratio of the poloidal to toroidal mean velocity that varies from $0.9$ to $0.4$ as the blade's curvature is changed from $72^{o}$ to $-72^{o}$ \cite{11}. In the blades and the impeller disk we impose either perfect ferromagnetic ($\vec{B} \times \vec{n} = \vec{0}$) or perfect conductor ($\vec{B} \cdot \vec{n} = 0$) boundary conditions ($\vec{n}$ is the surface unitary vector), null velocity and constant slope for the density ($\rho$) and pressure ($p$). For the wall at $Y = 4$ and at the other boundaries, the magnetic field is fixed to $10^{-3} T$ and oriented in the azimuthal $\vec{X}$ direction, mimicking an azimuthal disk magnetization as observed in the VKS experiment \cite{12}. The value of $10^{-3} T$ has been chosen to match the order of magnitude of the remnant magnetic field observed in the impeller, after a dynamo has been switched off. At the top boundary, the velocity is fixed to $\vec{V}=(10, 0, -10\Gamma) $ m/s, mimicking the impinging velocity field due to Ekman pumping towards the impeller.  The density is fixed to $931 kg/m^{3}$ in the left wall outside the blade ($X=0$) and has a constant slope in the rest. The pressure is calculated as $p = \rho v^{2}_{s}/\gamma $ with $\gamma = 5/3$ the specific heat ratio and $v_{s} = 250 m/s$ the sound speed. The $v_{s}$ value is one order smaller than the real sound speed in liquid Sodium to keep a time step large enough for the simulation to remain tractable. The consequence is a small enhancement of the compressible properties of the flow (subsonic low Mach number flow or pseudo-incompressibility regime), nevertheless the impact on the simulations is small and the incompressible nature of the liquid sodium flow is preserved.
  
The numbers of grid points are typically $128$ in the (X) and (Z) directions and $256$ in the (Y) direction for the simulations with kinetic Reynolds number $R_{n}=\rho V L/\eta=200$ with $L=1 \, m$ and $\eta$ the dynamic viscosity. For the simulations with $R_{n}=1000$ we double the resolution in each dimension. We chose to illustrate our results with the $R_{n}=200$ cases, which are more didactic to explain the physical mechanisms at play than the $R_{n}=1000$ cases, since in the latter the system enters a chaotic regime due to the precession of the whirl vortex, leading to a more complex regime to analyze. However, we demonstrate in the text and Table \ref{2} that the trends in both configurations are the same. The robustness of the model was further tested for different values of magnetic diffusivity ($\lambda$) as showed in Fig. 1C for a range of simulations with magnetic Reynolds number $R_{m}= V L/\lambda$ taking values between $10$ to $10^4$. The magnetic energy decreases when the magnetic diffusivity increases; in the simulation with $R_{m} = 10$ no enhancement of the magnetic field is observed, while for a simulation with $R_{m} = 100$ the magnetic energy is around 6 times smaller than the $R_{m} = 10^{4}$ case. The effective magnetic Reynolds number of the numerical magnetic diffusion due to the model resolution corresponds to $R_{m} \approx 6 \cdot 10^{3}$ in the simulation with $R_{n}= 200$. The simulations we show in this paper are performed using the numerical magnetic diffusion of the code for simplicity. We perform  hydrodynamic (HD) simulation with no magnetic field and  magnetohydrodynamic (MHD) simulations, with perfect ferromagnetic and perfect conductor boundary conditions at the blades and impeller disk.

\begin{figure}[h]
\includegraphics[width=0.8\columnwidth]{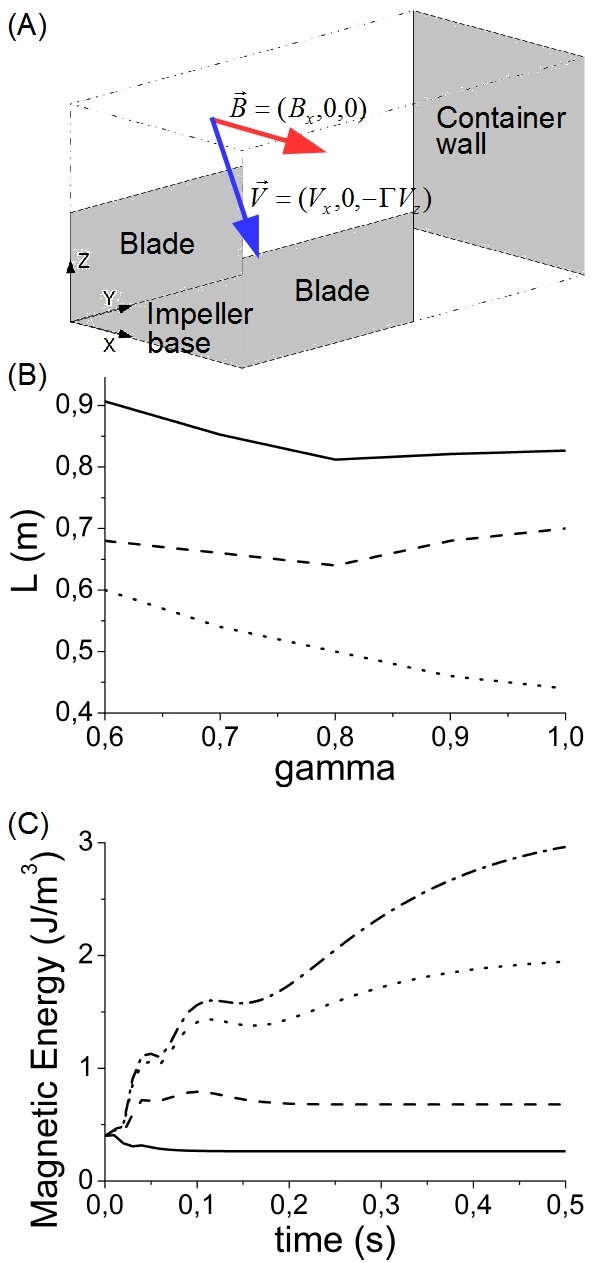} 
\caption{(A) Simulation domain for a portion of the flow in between two blades: X,Y and Z directions correspond to local azimuthal, radial and vertical direction with $X \in [0, 2], Y \in [0, 4], Z \in [0, 2] $.(B) Whirl location versus $\Gamma$ value. (C) Magnetic field for simulations with different magnetic diffusion.}
\label{1}
\end{figure}

\paragraph*{Vortex location and amplitude}
We have run several HD simulations with increasing Reynolds number $R_{n}$  and $\Gamma$ ranging from $0.6$ to $1$. In all cases, we observe that the impinging of the velocity at the impeller generates a radial helicoidal vortex, with a kinetic helicity increasing from 50 to \textcolor{red}{330 $ms^{-2}$ when $R_{n}$ is varied from $10$ to $1000$}, meaning that the vortex becomes more and more focused. A visualization of the resulting helicoidal vortex is displayed on Fig. 2 at $R_{n}=200$. At a given $R_n$, increasing the value of $\Gamma$ has two main effects: first, it increases the vortex kinetic helicity proportionally to $\Gamma$ (data not shown). Second, it changes the location of the vortex with respect to the blades (see Fig. 1B): as $\Gamma$ is increased, the vortex is pushed towards the upstream impeller, moving from $X=0.75 m$ for $\Gamma=0.6$ to $X=0.55 m$ for $\Gamma=1$, but it is in the configuration with $\Gamma=0.8$ where the whirl is closer to the left blade and the impeller base (see Fig. 1B, solid line). This intensification of the vortex and its displacement closer to the wall can potentially make it more sensitive to any current circulating within the blades. These results are compatible with HD simulations performed by other authors \cite{16}. To study this effect, we thus introduce a background magnetic field to investigate this interplay between the vortex flow and the magnetic field.

\paragraph*{Influence of boundary conditions}
We now fix $R_n=200$, $\Gamma=0.8$ (corresponding to TM 73 impeller rotating in the unscooping direction in the VKS experiment) and apply a large scale magnetic field in the azimuthal (X) direction at the boundaries, with ferromagnetic (denoted as Bvecn) or perfect conductor (denoted as Bdotn) boundary conditions at the impeller. In both cases, there is a generation of a magnetic field in the radial (Y) direction. The magnetic field lines are collimated by the vortex leading to a local enhancement of the magnetic field. In the (Bvecn) case (Fig. 2A) the magnetic field is almost 5 times larger than in the (Bdotn) case (Fig. 2B) pointing out the essential role played by the magnetic field boundary conditions in making the interaction between the flow and the magnetic field more efficient. An explanation of this can be found by investigating the geometry of the current streamlines (Fig. 3). The currents in the ferromagnetic walls are parallel to the surface (gray arrows) and are shorted out with the current lines inside the fluid, as can be observed for the current lines plotted nearby the upstream blade and the impeller base (black lines, Fig. 3A), not connected with the wall and creating a current whirl. This short-circuit avoids the transfer of magnetic energy from the fluid to the wall, \textcolor{red}{resulting in a tighter collimation and enhanced magnetic field amplitude overall}. For the conducting walls, the scenario is the opposite (Fig. 3B): the currents are perpendicular to the blades and the impeller base, connected with the current lines inside the fluid.  Therefore the fluid accumulates a smaller amount of magnetic energy resulting in a weaker magnetic field collimation and enhancement. To quantify this effect, we define the magnetic energy as a volume average in a region nearby the whirl $[ME] =\int ME d\vec{x}/\int d\vec{x}$, provided in Fig. 4D. The amount of magnetic energy in the ferromagnetic wall case is 7 times the one computed in the conducting wall case. In the sequel, we keep this notation for volume averaged quantities.

\begin{figure}[h]
\includegraphics[width=0.8\columnwidth]{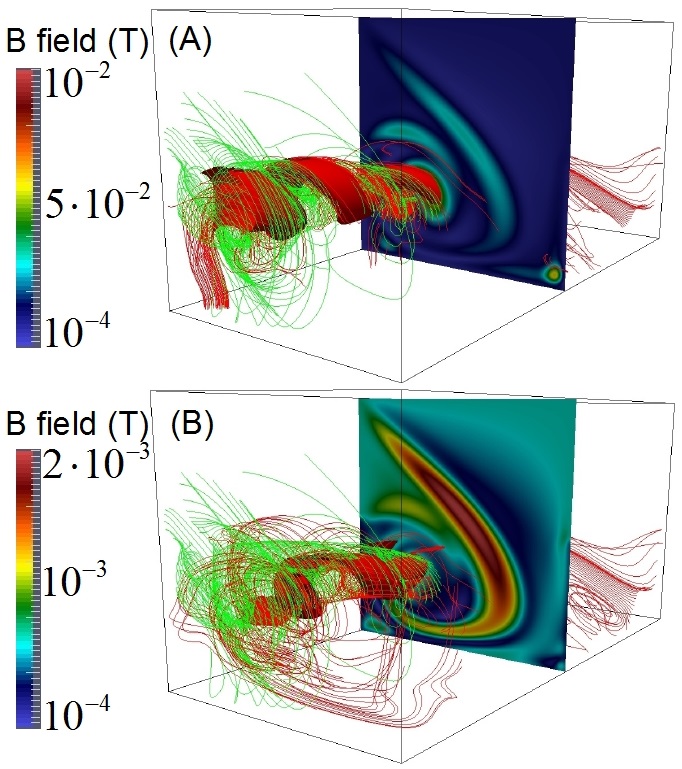} 
\caption{Magnetic and velocity field stream lines at $R_n=200$ and $\Gamma=0.8$ in the perfect ferromagnetic (A) and conductor (B) simulations in the all domain. Isocontour of the magnetic field (0.005 T in the ferromagnetic and 0.001 T in the conductor cases). Magnetic field module in the Y = 2 plane. }
\label{2}
\end{figure}

\begin{figure}[h]
\includegraphics[width=0.8\columnwidth]{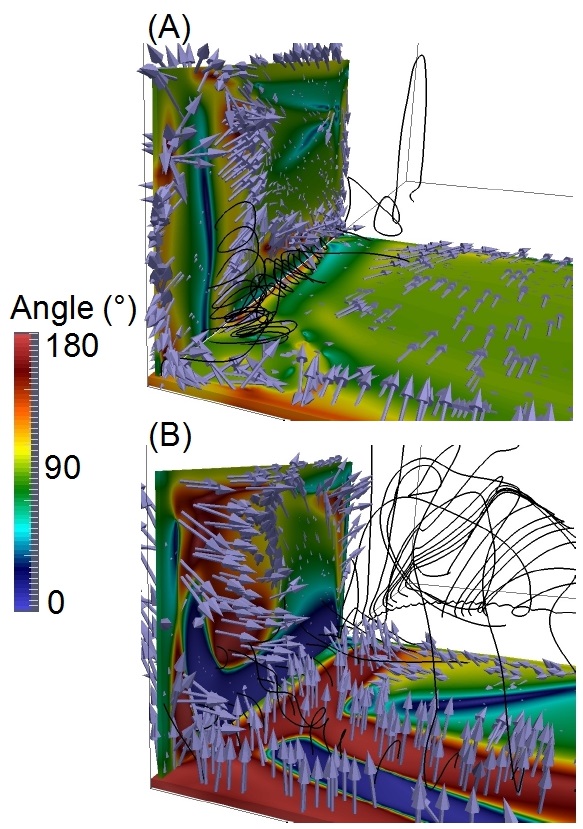} 
\caption{Angle between the current and the unitary surface vector in the Bvecn configuration (A) and the Bdotn configuration (B) in a zoomed box. The arrows indicate the orientation of the current on the blades and black lines are current stream lines nearby the upstream blade.}
\label{3}
\end{figure}

\paragraph*{Lorentz force feedback on vortex flow}
The magnetic field reaches locally values of $10^{-2}T$ (100 gauss) but there is no discernible influence on the mean flow helicity, as can be seen in Fig. 4A: the kinetic helicity time evolutions ($[\vec{v}\cdot\vec{\omega}]$, with $\vec{\omega}$ the vorticity) overlap for the Bvecn, Bdotn and HD simulations. Moreover, the current helicity ($[\vec{B}\cdot\vec{J}/\rho]$, with $\vec{J}$ the current density) is 5 times larger in the Bvecn simulation (Fig. 4B), but still 2 orders of magnitude smaller than the kinetic helicity. As a result, the total helicity ($He_T= [\vec{B}\cdot\vec{J}/\rho - \vec{v}\cdot\vec{\omega}]$) is dominated by the kinetic term, so that it is not influenced by the magnetic field, whatever the boundary conditions (Fig. 4E).  In contrast, the helicity of the fluctuations, defined as $He_f= [\vec{B'}\cdot\vec{J'}/\rho - \vec{v'}\cdot\vec{\omega'}]$ where the $'$ denotes the fluctuating part wrt the time-average, is sensitive to the boundary conditions (Fig. 4C and  4F). Splitting the fluctuating helicity into a current $ JH$ and a kinetic $KH$ parts, we see in Figure 4F that, in the Bdotn case, $ JH\ll KH$ while, in the Bvecn case, $JH$ dominates, showing that the magnetic field is strong enough to perturb the {\sl velocity fluctuations}.  

\begin{figure}[h]
\includegraphics[width=0.9\columnwidth]{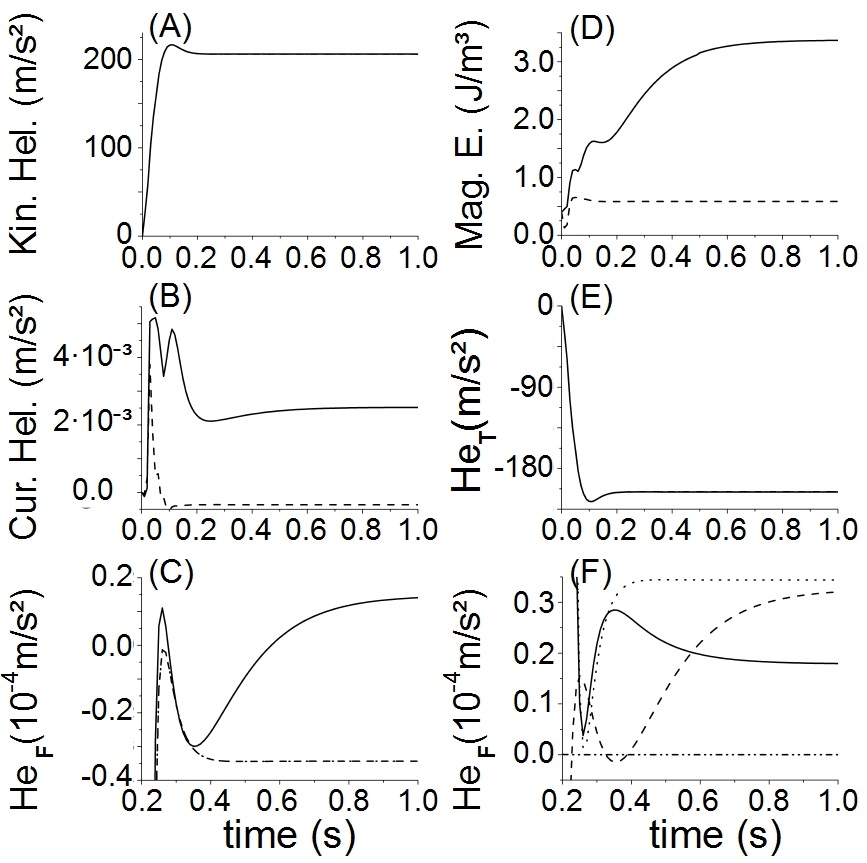} 
\caption{(A) Kinetic helicity. (B) Current helicity. (C) Magnetic energy. (D) Total helicity. (E) Helicity of fluctuations. 
(F) Kinetic and current helicity of fluctuations in the Bvecn and Bdotn cases. All values are averaged in a volume localized around the whirl.}
\label{4}
\end{figure}

To further quantify the variations of the fluctuating helicity for our 5 different setups (HD reference run, ferromagnetic and conductor cases for both $R_n$ number values), we have computed the 9 components of the helicity tensor $h_{ij}$ (data not shown), defined as  $h_{ij} = \epsilon_{ikn} \langle  u_{k}^{'} \partial_{j} u_{n}^{'} \rangle $. Due to the moderate Reynolds number in the $R_n=200$ simulations, the fluctuations are not very strong (of the order of 1 per cent of the mean velocity), so that the representative amplitudes of these quantities (maximum of the order of $10^{-3}$ m/s$^2$) are several orders smaller than in the $R_n=1000$ simulations. In order to analyze what an hypothetical $\alpha^{2}$ dynamo loop based on the components of the helicity tensor (e.g. $B_y\to^{h_{yy}, h_{zy}} B_x \to^{h_{xx}, h_{zx}} B_y$) would yield, one can define a gain factor $G$  using the ratio between the time averaged components of the dynamo loop for the ferromagnetic and conductor cases (e.g. $G = (\langle h_{ij} h_{km} \rangle)_{ferro} / (\langle h_{ij} h_{km} \rangle)_{cond}$). This allows us to assess what boundary condition configurations lead to the largest dynamo loop enhancement. In Table \ref{2}, we show first for the $R_n=200$ case that the gain factor is for most components greater than 1. This result confirms that the simulations with ferromagnetic boundary conditions are more efficient to collimate the flow and to induce the magnetic field, trend that is further reinforced for the simulations with $R_{n} = 1000$ where all components of the gain factor are now greater than 1.

\begin{table}[h]
\centering

\begin{tabular}{c}
$R_n=200$ \\
\end{tabular}

\begin{tabular}{c | c }
$ h_{xx} h_{zx} $ = 0.26 & $h_{yy} h_{xx} $ = 0.01 \\
$ h_{yy} h_{zx} $ = 1.90 & $ h_{yy} h_{zy} $ = 1.92  \\
$ h_{zy} h_{xx} $ = 0.01 & $ h_{zy} h_{zx} $ = 53.50 \\
\end{tabular}

\begin{tabular}{c}
$R_n=1000$ \\
\end{tabular}

\begin{tabular}{c | c }
$ h_{xx} h_{zx} $ = 1.90 & $h_{yy} h_{xx} $ = 2.33 \\
$ h_{yy} h_{zx} $ = 2.48 & $ h_{yy} h_{zy} $ = 1.43  \\
$ h_{zy} h_{xx} $ = 1.10 & $ h_{zy} h_{zx} $ = 1.17 \\
\end{tabular}

\caption{Gain factor (G) between the ferromagnetic and conductor cases for the simulations with $R_n=200$ and $R_n=1000$.}
\label{2}
\end{table}

\begin{figure}[h]
\includegraphics[width=0.8\columnwidth]{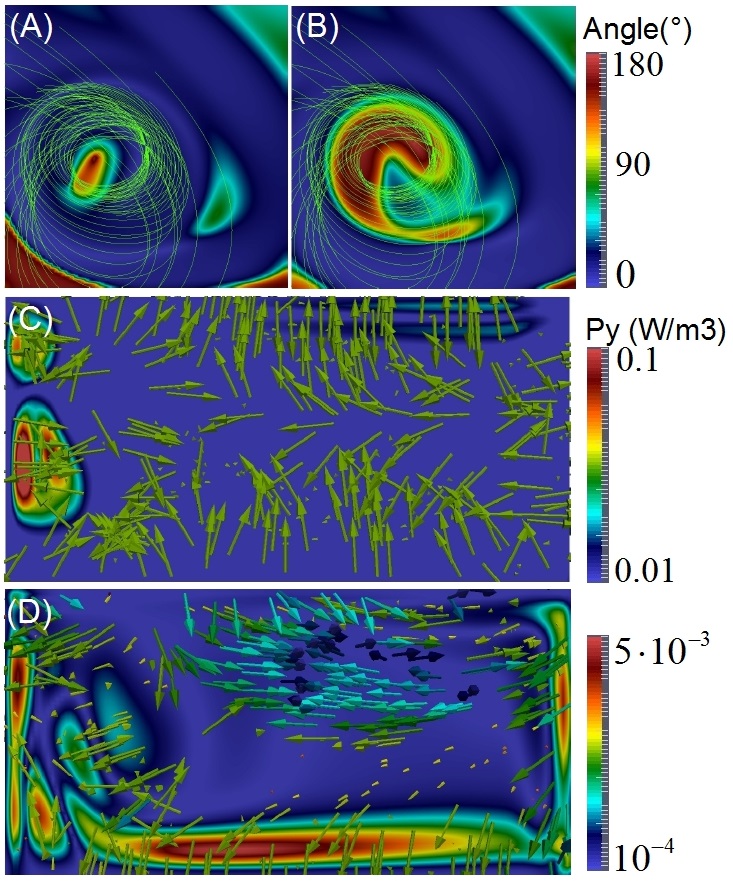} 
\caption{Angle between the velocity and the magnetic fields for (A)  the Bvecn and (B) the Bdotn simulation (plane $Y = 1$) at $t=1$ s.  Properties of the Poynting vector in the upstream blade: (C) Bvecn, (D) Bdotn cases. The arrows direction (resp. color) codes its orientation  (resp.  its angle with the unitary vector normal to the surface). The color in the blade codes its modulus.}
\label{5}
\end{figure}

Figures 5A-5B  show that regions of alignment or anti-alignment of  the magnetic and the velocity fields are associated with maxima of the magnetic energy. 
To clarify the magnetic energy transfer from the fluid to the boundary we calculate the Poynting vector on the upstream blade, defined as $ \vec{P_{y}}  = \frac{1}{\mu_{0}} \left(\vec{E} \times \vec{B} \right)$, with $\vec{E}$ the electric field. The Poynting vector in the Bvecn case is parallel to the surface so there is not a net flux of magnetic energy leaving the system, while in the Bdotn simulation the Poynting vector is not parallel to the surface so there is a net flux of magnetic energy. As a consequence, the magnetic energy in the Bvecn case remains in the fluid while in the Bdotn simulation it can be transferred to the wall.


\paragraph*{Discussion}
We have shown that the impinging of  the large scale flow upon the impeller generates a coherent radial vortex between the blades.  This flow collimates any existing magnetic field that in return modifies locally the velocity fluctuations and the helicity tensor. From Table I, we see that the products of the dynamo loop $h_{yy}h_{zx}$, $h_{yy}h_{zy}$ and $h_{zy}h_{zx}$ increase when changing the wall properties from conducting to ferromagnetic material in the $R_n=200$ simulations, trend that is even more robust in the non stationary $R_n=1000$ cases where all the products are larger for a ferromagnetic material. Within the first order smoothing approximation, the $\alpha$-tensor is defined as $\alpha=\tau h$ \cite{7}, where $\tau$ is the correlation time. This effect therefore leads to a \textcolor{red}{net} increase of the efficiency of the $\alpha^2$ dynamo mechanism, e.g. $B_y\to^{h_{yy}, h_{zy}} B_x \to^{h_{xx}, h_{zx}} B_y$ in the case of ferromagnetic boundary conditions as soon as the disks are magnetized. This effect is stronger when the vortex is closer to the wall, which is piloted by the parameter $\Gamma$. This effect may therefore explain why changing the blades orientation and the material of the impeller has an impact on the dynamo threshold  and validates the use of larger values of $\alpha$ when ferromagnetic blades are used, and the disks are magnetized \cite{13,14}. It has often been said the the VKS dynamo is just a rotating magnet because of the essential role of the soft iron. If it were the case, the dynamo properties would not be changed by the flow geometry, as it is observed. In the present work, we show that the soft iron may actually enhanced the dynamo cycle by retroaction of the magnetic field onto the velocity fluctuations responsible for the alpha effect, effect mediated by an helicoidal radial vortex that collimates the magnetic field. Even though the present scenario was elaborated at moderate values of the kinetic Reynolds number, we believe that it remains valid for the conditions realized in the experiments. The next step is to further increase the degree of turbulence and to push towards even more realistic values such as for instance lower magnetic Prandtl number. From an experimental point of view, it would be interesting to seek alternative ways to reinforce the $\alpha$-effect between the blades just by mechanics, by enhancing flow collimation rather than through the use of material favoring strong localized magnetic fields.

\begin{acknowledgments}
We have received funding by the Labex \textcolor{red}{ PALM/P2IO/LaPSIS} (VKStars grant number 2013-02711), INSU/PNST and ERC PoC grant 640997 Solar Predict. We thank Miki Cemeljic, Wietze Herreman and the VKS team  for fruitfull discussions.
\end{acknowledgments}

\end{document}